\documentclass[preprint2]{aastex}

\usepackage{amsmath}
\usepackage{bm}
\usepackage{kotex}

\slugcomment{Draft version Jun. 29, 2017}

\shorttitle{Balmer Wings in Z~And and AG~Dra}
\shortauthors{Chang et al.}

\begin{document}


\title{
Broad Wings around H$\alpha$ and H$\beta$  
in the Two  S-Type Symbiotic Stars Z~Andromedae and AG~Draconis
}


\author{Seok-Jun Chang\altaffilmark{1},  
 Hee-Won Lee\altaffilmark{1}, 
 Ho-Gyu Lee\altaffilmark{2}, 
 Narae Hwang\altaffilmark{2}, 
 Sang-Hyeon Ahn\altaffilmark{2}, 
 and Byeong-Gon Park\altaffilmark{2,3}}
\altaffiltext{1}{Department of Physics and Astronomy, Sejong University, Gwanjin-gu, Seoul 05006, Korea}
\altaffiltext{2}{Korea Astronomy and Space Science Institute, Daejon, 34055, Korea}
\altaffiltext{3}{Astronomy and Space Science Major, University of Science and Technology, 217, Gajeong-ro, Yuseong-gu, Daejeon, 34113, Korea}





\begin{abstract}

Symbiotic stars often exhibit broad
wings around Balmer emission lines, whose origin is still controversial. We present
the high resolution spectra of the S type symbiotic stars Z~Andromedae and AG~Draconis obtained with the ESPaDOnS
and the 3.6 m {\it Canada France Hawaii Telescope} to investigate the broad wings around H$\alpha$ and H$\beta$.
When H$\alpha$ and H$\beta$ lines are overplotted in the
Doppler space, it is noted that H$\alpha$ profiles are overall broader than H$\beta$ in these two objects.
Adopting a Monte Carlo approach, we consider the formation of broad wings of H$\alpha$ and H$\beta$
through Raman scattering of far UV radiation around Ly$\beta$ and Ly$\gamma$
and Thomson scattering by free electrons.
Raman scattering wings are simulated by choosing an \ion{H}{1} region with a
neutral hydrogen column density $N_{HI}$ and a covering factor $CF$.
For Thomson wings, the ionized scattering region is assumed to cover fully the Balmer emission nebula and
is characterized by the electron temperature $T_e$ and the electron column density $N_e$.
Thomson wings of H$\alpha$ and H$\beta$
have the same width that is proportional to $T_e^{1/2}$. However, Raman wings of
H$\alpha$ are overall three times wider than H$\beta$ counterparts, which is attributed to different cross
section for Ly$\beta$ and Ly$\gamma$. Normalized to have the same peak values and presented in the
Doppler factor space. H$\alpha$ wings of Z~And and AG~Dra are observed to be significantly wider than H$\beta$ counterpart,
favoring the Raman scattering origin of broad Balmer wings.

\end{abstract}





\section{Introduction\label{sec:intro}}

Symbiotic stars are wide binary systems of a hot white dwarf and a mass losing giant \citep[e.g.][]{kenyon09}.
They are important objects for studying the mass loss and mass transfer processes.
Their UV-optical spectra are characterized by prominent emission lines with a large range of ionization encompassing C~II and O~VI. 
 Symbiotic activities including outbursts and bipolar outflows 
are attributed to the gravitational capture of some fraction of slow 
stellar wind from the giant component by the hot white dwarf \citep[e.g.,][]{skopal08, mikolajewksa12}. 

According to IR spectra, symbiotic stars  are classified into `S' type
and `D' type. `D' type symbiotic stars exhibit an IR excess indicative of the presence of a warm dust component
with $T_D\sim 10^3{\rm\ K}$ \citep[e.g.][]{angeloni10}, whereas no such features are found in the spectra of
`S' type symbiotics. The orbital properties also differ greatly between these two types of symbiotics.
The orbital periods are found to be of order several hundred days for `S' type symbiotics, implying that the binary
separation is about several AU \citep[e.g.][]{mikolajewksa12}. However, the orbital periods of D type symbiotic stars are notoriously difficult to measure 
and only poorly known \citep[e.g.][]{schmid96, schmid00}.

Symbiotic stars show unique spectroscopic features at around
6825 \AA\ and 7082 \AA, which are formed through
Raman scattering of far UV resonance doublet \ion{O}{6}~$\lambda\lambda$1032 and 1038. 
\cite{schmid89} proposed that an \ion{O}{6}$~\lambda$1032 photon
is incident on a hydrogen atom in the ground state, which subsequently 
de-excites into $2s$ state emitting an optical photon at 6825 \AA. 
The 7082 feature is formed in a similar way when  
\ion{O}{6}~$\lambda$1038 photons are Raman scattered by neutral 
hydrogen atoms. The cross section for \ion{O}{6} is of order $10^{-22}{\rm\ cm^{2}}$ requiring the presence
of a thick \ion{H}{1} component in the vicinity of a hot emission nebula.

High resolution spectroscopy shows 
that most symbiotic stars exhibit broad wings around H$\alpha$, which often extend to several thousand ${\rm\ km\ s^{-1}}$ \citep[e.g.,][]{vanwinckel93, ivison94, skopal06, selvelli00}.
\cite{nussbaumer89} proposed that Raman scattering by atomic hydrogen can give rise to 
broad wings around Balmer emission lines. \cite{lee00} showed that the broad wings of many symbiotic stars are consistent with a profile proportional to $\Delta\lambda^{-2}$  that is expected of Raman scattering wings. Similar results for broad H$\alpha$ wings
in a number of planetary nebulae and symbiotic stars were obtained by \cite{arrieta03}.
These objects share the common property that a thick 
neutral component is present in the vicinity of a hot ionizing source.

Broad wings can also be formed through Thomson scattering, where emission line photons are scattered off
of fast moving electrons to get broadened \citep[e.g.][]{sekeras12}.
In particular, \cite{sekeras12} performed profile analyses of broad wings of \ion{O}{6} and \ion{He}{2} emission lines 
and estimated the Thomson scattering optical depth and electron temperature 
of the symbiotic stars, AG~Draconis, Z Andromedae and V1016~Cygni.

Z~And is regarded as a prototypical symbiotic star with an M-type giant whose mass is $\sim 2{\rm\ M_\odot}$ \citep[e.g.][]{murset99}.
The mass of the white dwarf has been reported to be $0.75{\rm\ M_\odot}$. The orbital period is not accurately
known and presumed to be in the range $759.0\pm1.9$ days \citep[e.g.,][]{fekel00}.
Having a K-type giant component, AG~Dra is known to be a yellow symbiotic star \citep[e.g.][]{murset99}. It has been reported that the masses 
of the white dwarf and the giant components of AG~Dra 
are $1{\rm\ M_\odot}$ and $0.5{\rm\ M_\odot}$, respectively. The known binary orbital period of
550 days implies that the binary separation is $\sim 0.3 {\rm\ AU}$ \citep[e.g.,][]{fekel00}.  

In this paper, we present broad wings around H$\alpha$ and
H$\beta$ from high resolution spectra of the `S' type symbiotic
stars Z~And and AG~Dra obtained with the ESPaDOnS and the 3.6 m 
{\it Canada-France-Hawaii Telescope} (CFHT). The observed wing profiles are compared
to the theoretical model wings obtained with a Monte Carlo technique. We propose that
Raman scattering is consistent with the two wing features whereas other
wing formation mechanisms present difficulty in reconciling with the observation.

\section{Spectroscopic Observation and Broad Balmer Wings}

\begin{figure*}
\centering
\includegraphics[scale=.95]{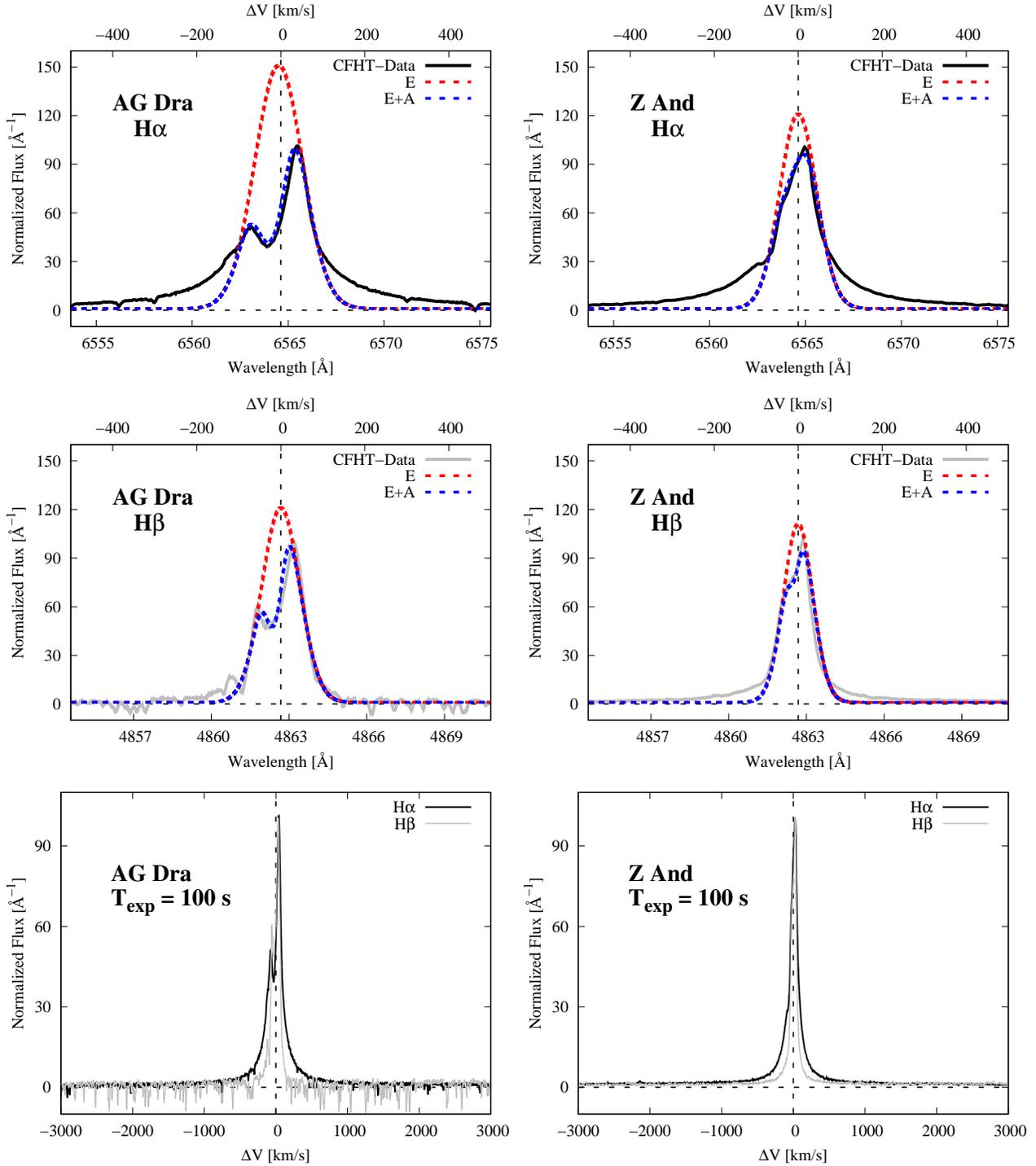}
\caption{
Spectra around  H$\alpha$ and H$\beta$ obtained with the high resolution spectrograph ESPaDOnS and
3.6 m {\it CFHT} of the two `S' type symbiotic stars AG~Draconis (left panel) and Z~And (right panel).
The upper and lower horizontal axes show the Doppler factor from line center in units of ${\rm km\ s^{-1}}$ and the wavelength in units of $\rm\AA$, respectively.
The top and middle panels show profiles of H$\alpha$ and H$\beta$ fitted with
a single Gaussian emission component and an emission component
with a Gaussian absorption, $G_E$ and $G_{EA}$.
The black dashed lines are $x$ and $y$ axes.
In the bottom panels, we overplot H$\alpha$ (black solid line) and H$\beta$ (grey solid line) in the Doppler factor space.
}
\label{cfht}
\end{figure*}

In Fig.~\ref{cfht}, we show the spectra around H$\alpha$ and H$\beta$ emission lines of Z~And and AG~Dra
obtained with the 3.6 m {\it CFHT} and ESPaDOnS on the nights of 2014 August 18 and September 6, respectively.
The two symbiotic stars were observed in the ``object only'' spectroscopic mode with the spectral resolution
of 81,000. The exposure time was 100 s. The standard procedure has been followed 
to reduce the observational data.

In the top and middle panels, the  lower horizontal axis shows the wavelength $\lambda$ and the upper horizontal axis shows 
the Doppler factor $\Delta V$ defined as
\begin{equation}
\Delta V = \left({{\lambda - \lambda_c} \over {\lambda_c}}\right) c,
\end{equation}
where $\lambda_c$ is the line center wavelength of either H$\alpha$ or H$\beta$
and $c$ is the speed of light. 
The weak underlying continuum is set to be a unit value and the profiles are normalized
so that the two emission lines have the same peak value of 101. 
In Appendix A, we discuss the normalization of continuum fluxes and profiles in this work and conversion into physical units.

In the bottom panels, we overplot the H$\alpha$ and H$\beta$ profiles 
in the Doppler factor space in order for quantitative comparisons of the profiles.
It is found that the H$\alpha$ profiles appear much broader than the H$\beta$ counterparts
in these two symbiotic stars. 
Furthermore, H$\alpha$ wings are more conspicuous than H$\beta$ wings
even when they are compared in the Doppler factor space.
Hereafter, for ease of profile comparisons, we describe the
profiles in the Doppler factor space.

 The top and middle panels of Fig.~\ref{cfht} show our fit  to H$\alpha$ and H$\beta$ 
using two functions, $G_E$ and $G_{EA}$
defined as follows
\begin{eqnarray}\label{core_para}
G_E &=& A_E\exp \left(-{{\Delta V}^2 \over {2 \sigma_E^2}} \right) \nonumber
\\
\tau_A &=& {\tau_{A0} \over {\sqrt{2\pi\sigma_A^2}} } \exp \left[-{({\Delta V - \Delta V_A})^2 \over {2 \sigma_A^2}} \right] 
\nonumber
\\
G_{EA} &=& G_E\exp(-\tau_A).
\nonumber
\\
\end{eqnarray}
Here, $G_E$ represents a single Gaussian emission profile characterized by the peak $A_E$ and width $\sigma_E$,
whereas $G_A$ represents an absorption component shifted by $\Delta V_A$ and characterized by
the peak optical depth $\tau_{A0}$ and the width $\sigma_A$. The function $G_{EA}$ describes a single Gaussian emission superimposed
by a single Gaussian absorption.

Table~\ref{core} shows the values of the parameters that characterize $G_E$ and $G_A$
for H$\alpha$ and H$\beta$ of the two objects.
The peaks and widths of H$\alpha$ emission are larger than those of
H$\beta$ emission, and also the same behavior is found for the H$\alpha$ and H$\beta$ absorption features.
In the case of AG~Dra, $\Delta V_A=-22 {\rm\ km\ s^{-1}}$ for both H$\alpha$ and H$\beta$ and
$\Delta V_A=-15 {\rm\ km\ s^{-1}}$ for Z~And.
The effect of the absorption profiles of H$\alpha$ and H$\beta$ can be seen in the shoulder parts
near $\Delta V \sim 100 \rm\ km\ s^{-1}$.
It should be noted that Balmer profiles of AG~Dra and Z~And differ significantly even in the core parts.
However, in this work, we focus on the wing profiles of these two symbiotic stars.

\begin{table*}[]
\centering
\caption{The parameters for core profiles of H$\alpha$ and H$\beta$ in Eq.~(\ref{core_para}) }
\label{core}
\begin{tabular}{|l|l|c|c|c|c|c|}
\hline
Object& Line& $A_E$ & $\sigma_E$ & $\tau_{A0}$ & $\sigma_A$ & $\Delta V_A$ \\ 
          &       &        & ${(\rm km\ s^{-1})}$ &     & $({\rm km\ s^{-1})}$ & ${(\rm km\ s^{-1})}$ \\\hline
AG~Dra&H$\alpha$ & 150  & 57  & 90  & 30  & -22 \\\hline
AG~Dra&H$\beta$  & 120  & 45  & 37  & 18  & -22  \\\hline
Z~And&H$\alpha$  & 120  & 44  & 20  & 27  & -15  \\\hline
Z~And&H$\beta$    & 110  & 35  & 10  & 13  & -15  \\\hline
\end{tabular}
\end{table*}

\section{Wing Formation and the Monte Carlo Approach} 

\subsection{Atomic Physics of Wing Formation}

In this work, we consider two physical mechanisms that may be responsible for formation of broad wings
around Balmer emission lines. One is Thomson scattering by free electrons and the other is Raman
scattering by atomic hydrogen.  In terms of the classical electron radius $r_e=e^2/(m_ec^2)=2.82\times 10^{-13}{\rm\ cm}$, 
the Thomson scattering cross section is given as
\begin{equation}
\sigma_{Th}={8\pi\over 3}r_e^2 = 0.665\times 10^{-24}{\rm\ cm^2}.
\end{equation}
The cross section $\sigma_{Th}$ is independent of the incident  photon wavelength, which  implies 
that line photons escape from the scattering region through
spatial diffusion in a way completely analogous to a random walk process. 

Thomson scattering is regarded as the low energy limit of Compton scattering, in which case there is
fractional energy exchange $\Delta E/E\sim h\nu/(m_ec^2)$. For a representative optical photon
with $h\nu=2{\rm\ eV}$ it amounts to $\Delta E/E \sim 4\times 10^{-6}$. This is equivalent
to a Doppler factor $\Delta V\sim 1.2{\rm\ km\ s^{-1}}$ or a redshift in the amount
of $0.024 {\rm\  \AA}$ for H$\alpha$ \citep[e.g.][]{hummer67}.

\cite{hummer62} presented the redistribution function in frequency space 
taking into account the non-coherency of Compton scattering. However, this effect
provides almost negligible contribution to broad Balmer wings having
widths in excess of $10^3 {\rm\ km\ s^{-1}}$. In view of this fact, previous works by
\cite{laor06} and \cite{kim07} took into consideration only thermal motions of free electrons
in order to to compute wing profiles of Thomson scattering.

In this work, we also consider only thermal motions of free electron in simulating Thomson scattering wings.
In the rest frame of a free electron, Thomson scattering 
is described as an elastic scattering process where
an incident photon changes its direction without any frequency shift. 
In the observer's frame, the scattered photon
acquires a Doppler shift corresponding to the electron velocity component along the difference in wavevectors
of incident and scattered radiation. 

We may expect that the width and strength of Thomson wings are mainly determined by the Thomson optical
depth $\tau_{Th}=N_e\sigma_{Th}$ and the electron temperature $T_e$, where $N_e$ is the electron 
column density given by the product of the electron density $n_e$ and the characteristic length associated
with the scattering region. The electron temperature representative of an emission nebula is $\sim 10^4{\rm\ K}$ \citep[]{osterbrock06},
for which the electron thermal velocity is
\begin{equation}
v_{th} =\sqrt{kT_e\over m_e} = 389\ T_4^{1/2} {\rm\ km\ s^{-1}},
\end{equation}
where $T_4 = T_e/10^4{\rm\ K}$. 
In terms of wavelength shift around H$\alpha$, this amounts to
$\Delta \lambda_{th} = v_{th}\lambda_{H\alpha}/c= 8.5{\rm\ \AA}$.

Broad wing features can develop around any spectral lines through Thomson scattering
due to wavelength independence of the cross section.
In constrast,
Raman scattering by atomic hydrogen can produce broad wings around hydrogen emission lines
except for Lyman lines, 
because the cross section is sharply peaked around hydrogen emission line centers.
Raman scattering is a generic term describing an inelastic scattering process involving  an incident photon
and an electron bound to an atom or a molecule.
In particular, far UV radiation near Ly$\beta$ and $Ly\gamma$ may be scattered by a hydrogen atom in the ground $1s$ state, 
which de-excites finally into the $2s$ state with the emission of an optical photon around H$\alpha$ and H$\beta$.
The energy conservation requires the relation between wavelengths $\lambda_i$ and
$\lambda_o$ of the  incident and scattered radiation, respectively,
given by
\begin{equation}
\lambda_o^{-1}=\lambda_i^{-1}-\lambda_{Ly\alpha}^{-1},
\label{raman_wvl}
\end{equation}
where $\lambda_{Ly\alpha}$ is the wavelength of Ly$\alpha$. 
A wavelength variation for the incident radiation
$\Delta\lambda_i$ corresponds to a much broader wavelength range $\Delta\lambda_o$ shown as
\begin{equation}
\Delta\lambda_o = \left({\lambda_o\over\lambda_i}\right)^2\Delta\lambda_i,
\end{equation}
from which we recognize the broadening factors $\lambda_o/\lambda_i\simeq 6.4$ for H$\alpha$ and 
$\lambda_o/\lambda_i\simeq 5.0$ for H$\beta$.

The interaction of a photon and an electron is described by the second order perturbation theory,
introduced in many textbooks on quantum mechanics including
\cite{sakurai67} and \cite{bethe57}. 
The cross section for Raman scattering is known as the Kramers-Heisenberg formula
\begin{eqnarray}
{d\sigma \over d\Omega} &=&r_e^2 
\left({\omega' \over \omega}\right)\Bigg|
{1\over m_e} \sum_I \left(
{({\bf p}\cdot \bm {\epsilon^{(\alpha')}})_{BI}
({\bf p}\cdot{\bm\epsilon^{(\alpha)}})_{IA}
\over E_I -E_A-\hbar\omega}
\right.
\nonumber \\
&+& \left.{({\bf p}\cdot{\bm\epsilon^{(\alpha)}})_{BI}
({\bf p}\cdot{\bm\epsilon^{(\alpha')}})_{IA}
\over E_I -E_A+\hbar\omega'}
\right) \Bigg|^2,
\end{eqnarray}
where ${\bm\epsilon}^{(\alpha)}$ and ${\bm\epsilon}^{(\alpha')}$ are polarization vectors associated
with the incident and scattered photons, respectively \citep[e.g.][]{saslow69,chang15}. 
Here, the initial state $A=1s$,  the final state $B=2s$
and the intermediate state $I$ covers all  bound $np$ states and continuum $n'p$ states.

The Wigner-Eckart theorem asserts that one can decompose the matrix elements into the radial part 
and the angular part, where the angular part is averaged over unpolarized incident radiation and summed over all possible
polarization states to yield the Rayleigh scattering phase function
\begin{equation}
< |{\bm\epsilon^{(\alpha)}}\cdot{\bm\epsilon^{(\alpha')}}|>={8\pi\over3}.
\end{equation} 
The matrix element associated with the momentum operator $p$ is simply related to the oscillator strength
by 
\begin{equation}
f_{n n'}={2\over m_e\hbar \omega_{nn'}}|p_{nn'}|^2,
\end{equation}
and the Kramers-Heisenberg formula can be rearranged in terms of the oscillator strengths \citep[e.g.,][]{nussbaumer89}.

As far as the broad wings around H$\alpha$ are concerned, the wavelength range of interest lies
in off-resonance regime sufficiently far
from the Ly$\beta$ line center compared to the thermal width.
Near Ly$\beta$, the cross section is dominantly contributed  by the $3p$ term, in which case the cross section
is excellently approximated by a Lorentzian function. According to \cite{lee13},
\begin{equation}
\sigma^{\rm Ram}(\omega) \simeq {5\over 128}\sigma_{Th}
\left({\omega_{32} \over \omega-\omega_{31}} \right)
 f_{13}f_{2s,3p},
\end{equation}
where the oscillator strengths are
\begin{equation}
f_{13}=0.0791, \quad
f_{2s,3p}=0.4349
\end{equation}
\citep[see e.g.][]{bethe57}.

\begin{figure*}
\centering
\includegraphics[scale=1.1]{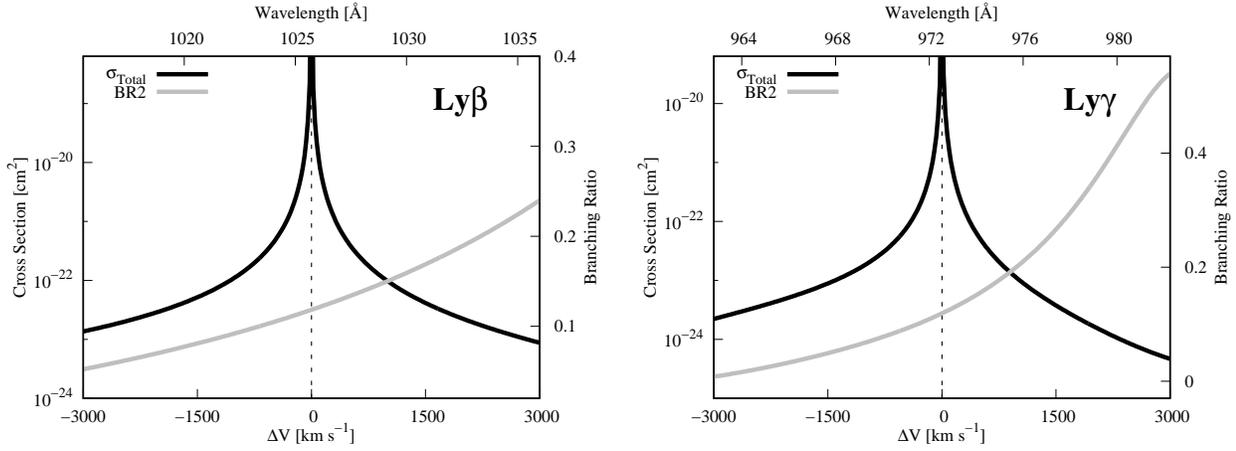}
\caption{Total cross section (black solid lines) and Balmer branching ratios (grey solid lines)  near Ly$\beta$ and Ly$\gamma$
computed using the Kramers-Heisenberg formula. 
}
\label{cross}
\end{figure*}

In Fig.~\ref{cross}, we present the total cross section near Ly$\beta$ and Ly$\gamma$,
which is characterized by the resonance behavior around line center. 
The figure also shows the branching ratio of Raman scattering into Balmer series, which is given
as the ratio of the Raman cross section de-excited to $2s$ state divided by the total cross section.
As \cite{chang15} pointed out, extremely broad Raman scattering wings of H$\alpha$ and H$\beta$
show asymmetry between redward and blueward.
Also, a hydrogen atom in the excited $3p$ level de-excites into the level $2s$ with a probability
of 0.12 and with the remaining probability of 0.88 it de-excites into the ground level.

\subsection{Scattering Geometries for Thomson and Raman Wings}
\begin{figure*}
\centering
\includegraphics[scale=0.9]{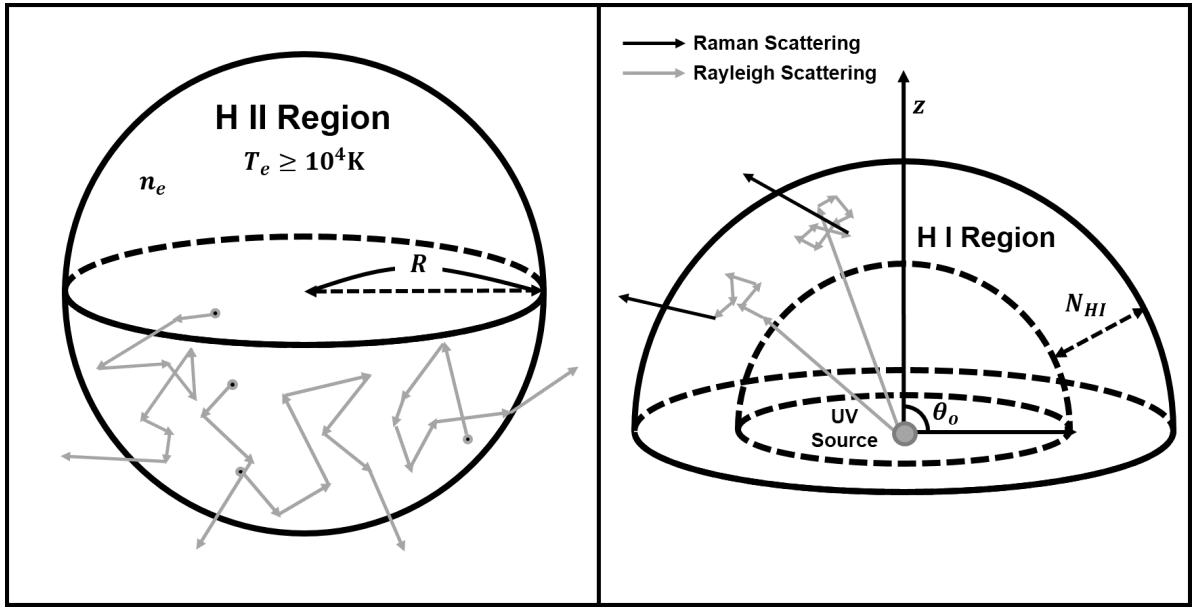}
\caption{Scattering geometries of the Thomson scattering (left panel) and Raman scattering (right panel) adopted in this work. In the case of Thomson scattering,
free electrons are distributed with a constant number density $n_e$ in a sphere with radius $R$. The sphere is 
characterized by the Thomson optical depth $\tau_{Th}=n_e\sigma_{Th}R$. H$\alpha$ line photons are
assumed to be generated uniformly inside the sphere.  In the case of Raman scattering, the scattering neutral region
takes a form of a partial spherical shell with a half-opening angle $\theta_o$. The far UV emission source coincides
with the center of the partial spherical shell. 
}
\label{scheme}
\end{figure*}

The left panel of Fig.~\ref{scheme} illustrates schematically the Thomson scattering geometry considered in this work.
We adopt a static \ion{H}{2} region in the form of a sphere with radius $R$ and a uniform electron number density $n_e$, where
hydrogen Balmer line photons are generated in a uniform fashion and Thomson scattered. 
We introduce the Thomson scattering optical depth 
$\tau_{Th} = n_e\sigma_{Th}R$ to characterize the \ion{H}{2} region.

The parameter space considered in this work is described by the electron temperature $T_e \geq 10^4\,\rm K$ and
the Thomson optical depth $\tau_{Th} \leq 0.1$.
The Monte Carlo simulation for Thomson wings starts with a generation of a Balmer line photon in a random place
inside the spherical \ion{H}{2} region with an initial frequency chosen to be a Gaussian random deviate with the standard
deviation of $\sigma_v=30 {\rm\ km\ s^{-1}}$ in the Doppler space. 
It should be noted that this velocity is quite smaller than the electron thermal velocity 
$v_{th} \sim 390 {\rm\ km\ s^{-1}}$ at $T_e = 10^4\, \rm K$.

We show a scattering geometry of Raman scattering wings in the right panel of 
Fig.~\ref{scheme}. \cite{dumm99} investigated  an eclipsing symbiotic star SY~Muscae to measure the \ion{H}{1} column density as a function 
of binary orbital phase. They found that $N_{HI}$ varies from $10^{21}{\rm\ cm^{-2}}$ to $10^{25}{\rm\ cm^{-2}}$,
where an extremely high column density of $10^{25}{\rm\ cm^{-2}}$
is limited to the orbital phase near $\phi_{orbit}=1$ with the giant companion in front of the white dwarf component. From this result, we may imagine that 
a neutral region with $N_{HI}\sim 10^{21}{\rm\ cm^{-2}}$ may surround the  binary system quite extensively. In this work, 
the scattering neutral region is assumed to be a partial spherical 
shell with a half-opening angle $\theta_o$ which is converted to the covering factor 
$CF=(1-\cos\theta_o)/2$. 

We locate the far UV emission source at the center 
of the partial spherical shell. Treating as a point source, we consider flat and isotropic 
continuum around Lyman series. The partial spherical \ion{H}{1} region is characterized 
by a neutral hydrogen column density $N_{HI}$ along the radial direction. 

\subsection{Monte Carlo Approach}

We adopt a Monte Carlo technique to produce simulated wings around H$\alpha$ 
formed through Raman scattering and Thomson scattering. We generate a photon in the emission region with an
initial wavevector ${\bf \hat k}$ and follow its path until escape from the scattering region.
Thomson scattering is characterized by the scattering phase function 
given by
\begin{equation}
f(\mu) \propto 1+\mu^2,
\end{equation}
where $\mu=\cos\theta_s$ is the cosine of the angle $\theta_s$ between the incident
and scattered wavevectors. The same phase function also describes Rayleigh and Raman scattering
as long as the incident photon is off-resonant where the deviation from line center is sufficiently larger than  
the fine structure splitting \citep{stenflo80}.

A convenient formalism incorporating the polarization information is the density matrix approach,
in which a Hermitian $2\times2$ matrix [$\rho$] is associated with a given photon. The density matrix elements
are easily interpreted in terms of the Stokes parameters $I, Q, U$ and $V$, where
\begin{eqnarray}
I &=& \rho_{11}+\rho_{22}
\nonumber \\
Q &=& \rho_{11}-\rho_{22}
\nonumber \\
U &=&{1\over2} {Re} [\rho_{12}]={1\over2}{Re}[\rho_{21}]
\nonumber \\
V &=& {1\over2}{Im} [\rho_{12}]=-{1\over2}{Im}[\rho_{21}].
\end{eqnarray}

For a photon propagating in the direction ${\bf\hat k}=
(\sin\theta\cos\phi, \sin\theta\sin\phi, \cos\theta)$, we may choose a set 
$\{  |{\bm\epsilon_1}>,|{\bm\epsilon_2}>  \}$ of basis vectors 
representing the polarization states, where
\begin{eqnarray}
|{\bm\epsilon_1}> &=& (-\sin\phi,\cos\phi,0), 
\nonumber \\
|{\bm\epsilon_2}> &=&  (\cos\theta\cos\phi, \cos\theta\sin\phi,-\sin\theta) .
\end{eqnarray}
The density matrix elements $\rho_{ij}'$ associated with the scattered radiation with ${\bf\hat k}_f
=(\sin\theta'\cos\phi',\sin\theta'\sin\phi',\cos\theta')$ are related to those for
the incident photon 
\begin{equation}
\rho_{ij}'=\sum_{kl}
<{\bf\hat e}_i' | {\bf\hat e}_k> \rho_{kl}
<{\bf\hat e}_j' | {\bf\hat e}_l>,
\end{equation}
where  $<{\bf\hat e}_j' | {\bf\hat e}_l>$ is an inner product between the two polarization basis vectors
$| {\bf\hat e}_l>$ and $| {\bf\hat e}_i'>$ associated with the incident and scattered photons, respectively.
A more detailed explanation can be found in  
\citep[e.g.][]{lee94, ahn15}.

In the Monte Carlo simulations, each scattering optical depth $\tau$ traversed by a 
photon in the scattering region is given in a probabilistic way as
\begin{equation}
\tau = -\ln X,
\end{equation}
where $X$ is a uniform random number between 0 and 1.
This optical depth $\tau$ is inverted
to a physical distance in order to locate the next scattering site. 
In the case of Thomson scattering, the physical distance $l_{Th}$ is given by
\begin{equation}
l_{\rm Th} = {\tau \over {n_e\sigma_{Th}}} .
\end{equation}

Using this value, we determine the next scattering position $\bm r_s$ by the relation
\begin{equation}
\bm r_s = \bm r_i + l_{\rm Th} \hat{\bm k},
\end{equation}
where $\bm r_i$ is the position vector of the previous scattering site. If ${\bm r_s}$ is outside the ionized nebular region,
the photon is considered to escape from the region to reach the observer. Otherwise,  
we iterate the process to find a new scattering position treating ${\bm r_s}$ as starting position.

In the case of transfer of a far UV photon near Ly$\beta$ and Ly$\gamma$,  the photon under study may be Rayleigh scattered several times before it becomes 
an optical photon through Raman scattering. The \ion{H}{1} region is assumed to be optically thin for optical photons near H$\alpha$ and H$\beta$ so that
a Raman scattered photon directly escapes to reach the observer.

The physical distance $l_{\rm Ram}$ corresponding to a probabilistic scattering optical depth $\tau$, 
\begin{equation}
l_{\rm Ram} = {\tau \over {n_{HI}\sigma_{tot}}},
\end{equation}
and the next scattering site is
\begin{equation}
\bm r_s = \bm r_i + l_{\rm Ram} \hat{\bm k}.
\end{equation}
At the location of ${\bm r_s}$, we generate a new wavevector ${\bm k_o}$. 
Using another uniform random number $X'$ and the Raman branching ratio $b$,  decision is made for the scattering type. If it  is Raman, then
the photon is regarded as an optical photon, which directly escapes in the direction of ${\bm k_o}$. Otherwise, the scattering
is Rayleigh and we determine the next scattering site by obtaining a revised $l_{\rm Ram}$.

In dealing with Rayleigh scattering, we fix the wavelength of a given photon 
because the thermal motion of an atomic hydrogen in H~I region is negligible compared to
the wing widths $> 10^3\ \rm km\ s^{-1}$.

On the other hand, Thomson scattered photons suffer diffusion in frequency space with a step corresponding 
to the thermal motion of free electrons.
The velocity of a free electron $\bm v_{th}$ is selected  in a probabilistic way using Gaussian random deviates.
The Thomson scattering is treated as elastic in the rest frame of the scatterer. Transforming to the observer's frame,
we obtain the relation between the wavelengths before and after scattering $\lambda_i$ and $\lambda_f$  
\begin{equation}
\lambda_f = \left(1-{{\hat{\bm k_f} \cdot \bm v_{th}}\over c}\right)
\left(1+{{\hat{\bm k} \cdot \bm v_{th}}\over c}\right)\lambda_i .
\end{equation}

\section{Simulated Broad Balmer Wing Profiles}

\begin{figure*}
\includegraphics[scale=0.85]{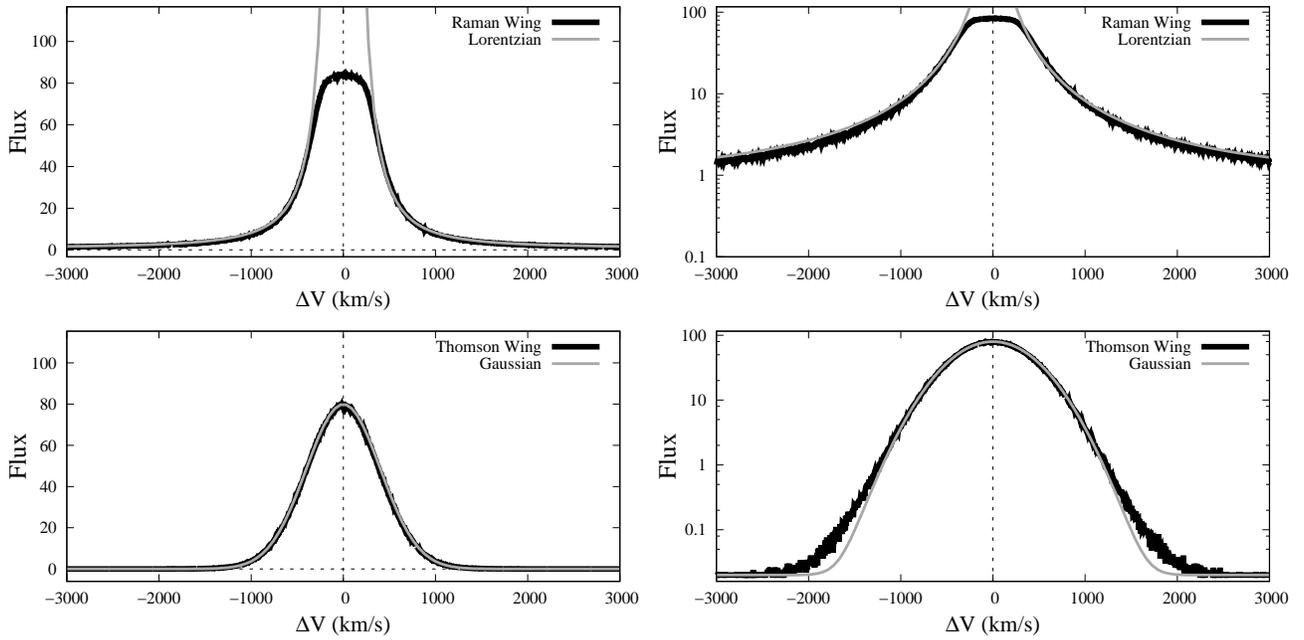}
\caption{Black solid lines and grey solid lines show our Monte Carlo results and fitting functions for broad wings. 
The top panels and the bottom panels show Raman and Thomson scattering wings formed in scattering geometries illustrated in the right and left  panels
Fig~\ref{scheme}, respectively. The fitting functions shown in grey solid lines are a Gaussian and a Lorentzian
for Raman wings and Thomson wings, respectively.
} 
\label{wings} 
\end{figure*}

In this section, we present broad wings obtained from our Monte Carlo calculations for Raman scattering with atomic hydrogen 
and Thomson scattering with free electrons.
Two representative wing profiles are shown in Fig.~\ref{wings}, where the top and bottom panels are for Raman 
and Thomson wings, respectively. The black thick solid lines show our Monte Carlo result and grey thin solid lines
are fitting functions.

In the left panels, the vertical scale is linear whereas
the right panels are presented in logarithmic scale in order for easy comparisons with fitting functions.  
In the bottom right panel, the Thomson wings show excess at $|\Delta V|> 1000 {\rm\ km\ s^{-1}}$
compared to the fitting Gaussian function.  Multiple Thomson scattering may result in deviation of the wing profile 
from a pure Gaussian, because Thomson scattering involves diffusion in frequency space as well as spatial
diffusion \citep[e.g.,][]{kim07}.

The Raman wing feature is fitted with a Lorentzian function, because 
the Raman cross section near Ly$\beta$ is roughly approximated by a Lorentzian function 
due to the proximity of resonance \citep[e.g.,][]{lee13}. The Thomson wing profile is fitted with a Gaussian function 
because the free electron region is assumed to follow a Maxwell-Boltzmann 
distribution. As $\Delta V$ is increased, the Gaussian nature forces Thomson wings to decay much more rapidly
than Raman wings.

\subsection{Thomson Wings}

\begin{figure*}
\centering
\includegraphics[scale=0.85]{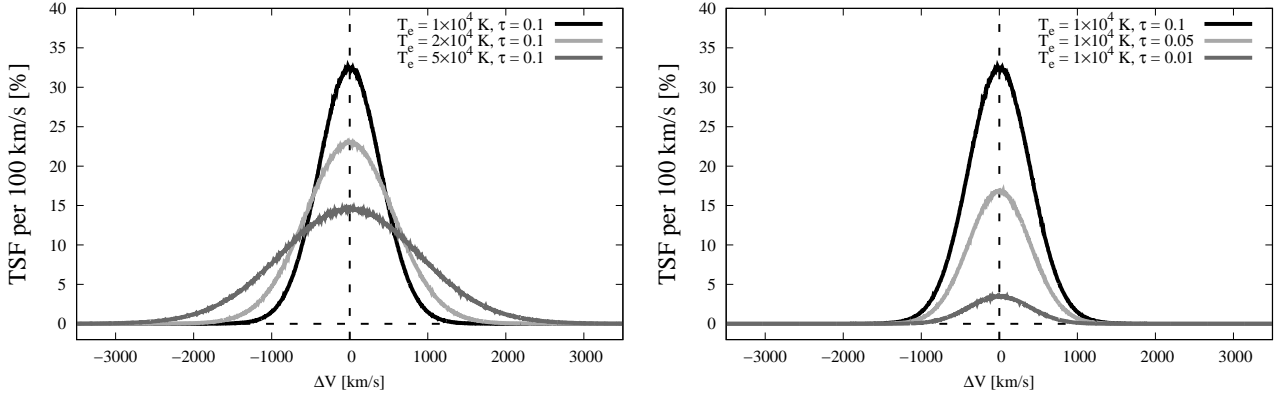}
\caption{Thomson wing profiles obtained from our Monte Carlo simulations for various values of electron temperature $T_e$
and Thomson scattering optical depth $\tau_{Th}$. The left panel shows Thomson wings 
for three values of $T_e =1\times 10^4, 2\times 10^4$ and $3\times 10^4{\rm\ K}$ with the Thomson optical depth fixed to $\tau_{Th}=0.1$. 
The right panel shows Thomson wings
for fixed $T_e=10^4{\rm\ K}$ and four values of $\tau_{Th}=0.01$, 0.05 and 0.1.
}
\label{thomson_wing}
\end{figure*}

Fig.~\ref{thomson_wing} shows Thomson wings obtained from our Monte Carlo simulations 
for various values of the electron temperature $T_e$ and Thomson optical depth $\tau_{Th}$.
The horizontal axis is the Doppler factor and the vertical axis is relative photon number represented in
linear scale.
In the Monte Carlo simulations, we collect only those photons Thomson scattered at least once
ignoring emergent photons without being scattered at all.

In order to describe the Thomson wings in a quantitative way, 
we introduce a parameter 'TSF (Thomson scattered fraction)' defined by
\begin{equation}
{\rm TSF}(\Delta V)d(\Delta V) = {N_{scat}(\Delta V) \over N_{init}} d(\Delta V).
\end{equation}
Here, $N_{init}$ is the number of photons initially emitted from the uniform \ion{H}{2} region
and $N_{scat}(\Delta V)d(\Delta V)$ represents the number of collected photons having a Doppler factor
in the interval $\Delta V$ and $\Delta V+d(\Delta V)$. For simplicity, we take
$d(\Delta V)=100\,\rm km\, s^{-1}$.
In Fig.~\ref{thomson_wing}, the vertical axis shows 'TSF'.

The left panels of Fig.~\ref{thomson_wing} show Thomson wings for $T_4 = 1 ,2$ and $5$
with the Thomson optical depth fixed to $\tau_{Th}=0.1$. Because of small Thomson optical depth,
the number of scattering averaged over all collected photons is slightly larger than unity. 
In this situation, the Thomson wings follow approximately  a Maxwell-Boltzmann distribution with a
width proportional to ${T_e}^{1/2}$. This is illustrated by the result that the total wing flux is almost
constant while the width increases as $T_e$ increases.

The right panels show Thomson wings for various values of the Thomson optical depth 
$\tau_{Th}= 0.01,0.05$ and $0.1$ at fixed $T_4=1$.
For $\tau_{Th}<1$, the wing flux is roughly proportional to $\tau_{Th}$ while the profile width remains unchanged.

\subsection{Raman Wings}

\begin{figure*}
\centering
\includegraphics[scale=0.58]{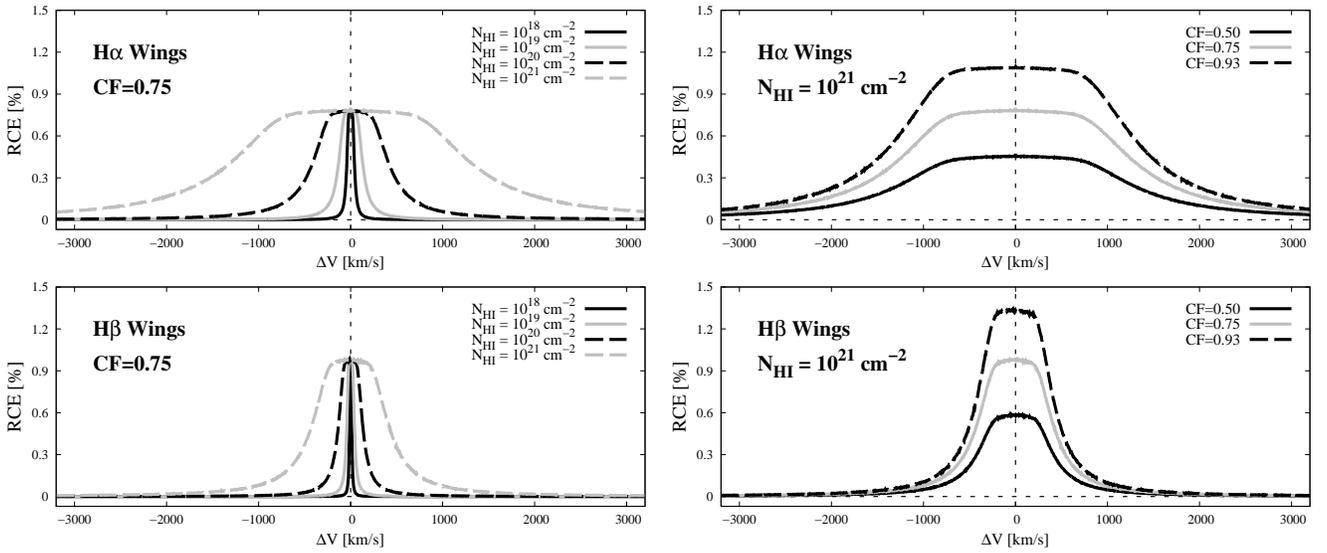}
\caption{Monte Carlo simulated Raman wings around H$\alpha$ (top panels) and H$\beta$ (bottom panels)
formed in a neutral partial spherical shell illustrated in the right panel of Fig.~\ref{scheme}. The left panels show Raman wings for  a fixed value of 
$CF=0.75$ varying the neutral hydrogen column density $N_{HI}$ in the range $10^{18}-10^{21}{\rm\ cm^{-2}}$. 
The right panels are Raman wings for various values of $\theta_o$ and a fixed $N_{HI}=10^{21}{\rm\ cm^{-2}}$.
}
\label{raman_wing}
\end{figure*}

Fig.~\ref{raman_wing} shows Raman wings near H$\alpha$ and H$\beta$ obtained with our Monte Carlo code,
 which simulates Raman wing formation in a neutral partial spherical shell illuminated by a flat continuum around Ly$\beta$ and Ly$\gamma$. 
For a flat incident continuum, the Raman wing profile is identified with
the Raman conversion efficiency (RCE) per unit scattered wavelength $\lambda_o$, defined by
\begin{equation}
{\rm RCE}(\lambda_o)d\lambda_o  = {N_{Raman} \over N_{total}}d\lambda_i.
\end{equation}
Here, $N_{total}$ is the number of incident photons from the flat and isotropic source 
per incident wavelength $\lambda_i$. The wavelength of Raman scattered radiation $\lambda_o$ is related to $\lambda_i$
by Eq.~\ref{raman_wvl}. We also denote by
$N_{Raman}$  the number of Raman scattered photons collected over all lines of sight. In Fig.~\ref{raman_wing},
the vertical axis shows 'RCE' as a function of the Doppler factor in the observed frequency space.
We consider various values of $N_{HI} = 10^{18}-10^{21}\rm\, cm^{-2}$ and $CF=0.5-0.93$.

Quantitatively we obtain ${\rm RCE} \sim 0.01$ for H$\alpha$ and H$\beta$ wings, respectively.
Quite unlike the case of Thomson wings, H$\alpha$ wings are much broader than those of H$\beta$, which is explained by
overall larger cross section around Ly$\beta$ than Ly$\gamma$. 
As is shown in Fig.~\ref{raman_wing}, the widths of Raman H$\alpha$ wings
are almost three times larger than those of H$\beta$ in the Doppler space.
The small Raman conversion efficiency is attributed to the smallness of the factor 
\begin{equation}
A={d\lambda_i \over d\lambda_o}=\left({\lambda_i \over \lambda_o}\right)^2,
\end{equation}
which gives the ratio of the one-dimensional volumes of incident frequency space and Raman frequency space.
Approximately the ratios are approximately $0.025$ and $0.04$ 
 near H$\alpha$ and  H$\beta$, respectively. 
A close look at Fig.~\ref{raman_wing} reveals higher 'RCE' for H$\beta$ than H$\alpha$ near line center, where
Raman scattering cross section is high. This is explained by the larger value of '$A$' factor for H$\beta$ than that for H$\alpha$.

Chang et al. (2015) showed that Raman wings around H$\alpha$, H$\beta$ and 
Pa$\alpha$ exhibit different widths and strengths when they are formed in a region with high column density 
$N_{HI} \geq 10^{22} \, \rm cm^{-2}$ due to significant differences in the cross section $\sigma^{Ram}(\omega)$ and branching channels.
The widths of Raman wings are roughly proportional to $N_{HI}^{1/2}$ as \cite{chang15} showed.
In the same work, the Raman conversion efficiency is saturated to remain flat near the line center, 
where the Raman scattering optical depth exceeds unity.

The right panels in Fig.~\ref{raman_wing} show our Monte Carlo results for three values of the covering factor '$CF$' at fixed $N_{HI}$.
In the case of high $CF$, 
Rayleigh escaping far UV photons may reenter the scattering region to provide an additional contribution to the final 'RCE'.
The effect of reentry is considered in this work, which results in nonlinear response of Raman wing flux
as a function of '$CF$'.

\subsection{Comparison of Simulated Raman and Thomson Wings}

\begin{figure*}
\centering
\includegraphics[scale=0.57]{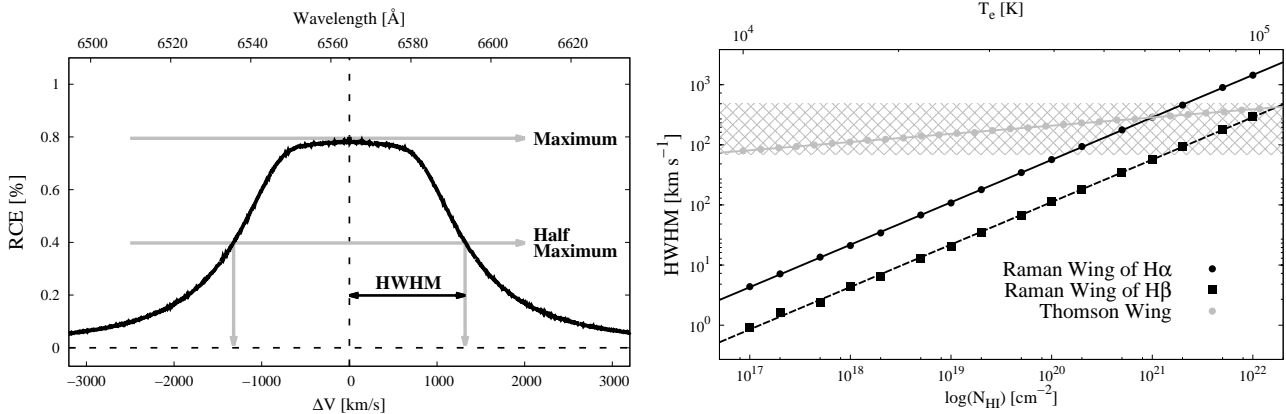}
\caption{Half width at half maximum (HWHM) of Thomson and Raman wings computed with a Monte Carlo technique.
HWHM of Thomson wings is presented with grey dots as a function of $T_e$ shown in the upper horizontal axis with a fixed $\tau_{Th}=0.1$.
Data for Raman H$\alpha$ and H$\beta$ wings are shown by black circle and square marks, respectively, as a function of $N_{HI}$
corresponding to the lower horizontal axis. 
}
\label{width}
\end{figure*}

In this subsection, we compare 
Raman wings around H$\alpha$ and H$\beta$ and Thomson wings produced in our Monte Carlo simulations.
Fig.~\ref{width} shows the HWHM of Raman wings around H$\alpha$ (black dots) and H$\beta$ (black squares)
obtained for various values of $N_{HI}=10^{17}-10^{22}\,\rm cm^{-2}$. The lower horizontal axis shows 
$N_{HI}$ in logarithmic scale. Note that the vertical scale is also logarithmic. We also plot
the HWHM of Thomson wings with grey dots for various values of the electron temperature 
$T_4 = 1-10$. The upper horizontal axis represents $T_e$ in logarithmic scale.
 It should be noted that in the Doppler factor space
Thomson wings of H$\alpha$ are identical with those of H$\beta$ due to the independence of Thomson cross
section on wavelength.

The linear relations shown in logarithmic scales in Fig.~\ref{width} imply that the HWHMs of Raman and Thomson wings are related to
$N_{HI}$ or $T_e$ by a power law.
We provide the fitting relations as follows
\begin{eqnarray}\label{eqn:relation}
{\rm HWHM}_{\rm Raman \, H\alpha} &=& 418\ N_{20}^{0.49} \rm\ km\ s^{-1}
\nonumber \\
{\rm HWHM}_{\rm Raman \, H\beta} &=& 135\ N_{20}^{0.49} \rm\ km\ s^{-1}
\nonumber \\
{\rm HWHM}_{\rm Thomson \;\;\,} &=& 532\ T_4^{0.49}  \rm\ km\ s^{-1},
\end{eqnarray}
where $N_{20}=N_{HI}/[10^{20}\ \rm cm^{-2}]$ is the \ion{H}{1} column density divided by $10^{20}{\rm\ cm^{-2}}$. 

The Raman wing widths are approximately proportional to $N_{HI}^{1/2}$, which is attributed to
the fact that the total scattering cross sections around Lyman series are approximately proportional to
$\Delta\lambda^{-2}$. On the other hand, the Thomson wing widths are proportional
to $T_e^{1/2}$, which reflects the fact that the electron thermal speed $v_{th}\propto T_e^{1/2}$.

From  Eq.~(\ref{eqn:relation}), one may notice that the 
HWHMs of Raman H$\alpha$ wings are approximately three times as large as those for H$\beta$ .
In the range of electron temperature $T_4=1-10$, the HWHM of Thomson wings lies in the narrow range
of $1,080\pm 550{\rm\ km\ s^{-1}}$, reflecting the fact that the width of Thomson wings is mainly determined
by the thermal electron speed $v_{th}$. 
Based on this result, it may be argued that Thomson scattering is inadequate to explain broad wings observed to exhibit  
$\rm{HWHM}<530 {\rm\ km\ s^{-1}}$, noting that an ionized region is hardly colder than $10^4{\rm\ K}$
\citep[e.g.][]{osterbrock06}. 
The same argument can be made for wings that may exceed several $10^3{\rm\ km\ s^{-1}}$
often reported in the planetary nebula M2-9, for which \cite{balick89} reported the presence of broad H$\alpha$
wings with a width $>10^3{\rm\ km\ s^{-1}}$.

For $N_{HI}=10^{20-21}{\rm\ cm^{-2}}$ and $T_e=10^{4-5}\ \rm K$, the widths of Raman H$\alpha$ wings and Thomson wings are comparable.  
In the case of Raman wings, H$\alpha$ wings are much broader than H$\beta$
counterparts by a factor of three, as stated earlier. The observational clear signature for Raman origin
is the unambiguous disparity of the wing widths of H$\alpha$ and H$\beta$. However, observationally H$\beta$ 
is weaker than H$\alpha$ rendering H$\beta$ wings often poorly measured.  
In contrast to simulations, no unequivocal isolation of
wing photons or scattered photons from the entire Balmer emission flux is apparent with the observed spectra, preventing one from
measuring the HWHMs of the observed Balmer wings.

\subsection{Comparison with the Observed Spectra of AG~Dra and Z~And}

\begin{figure*}
\centering
\includegraphics[scale=0.86]{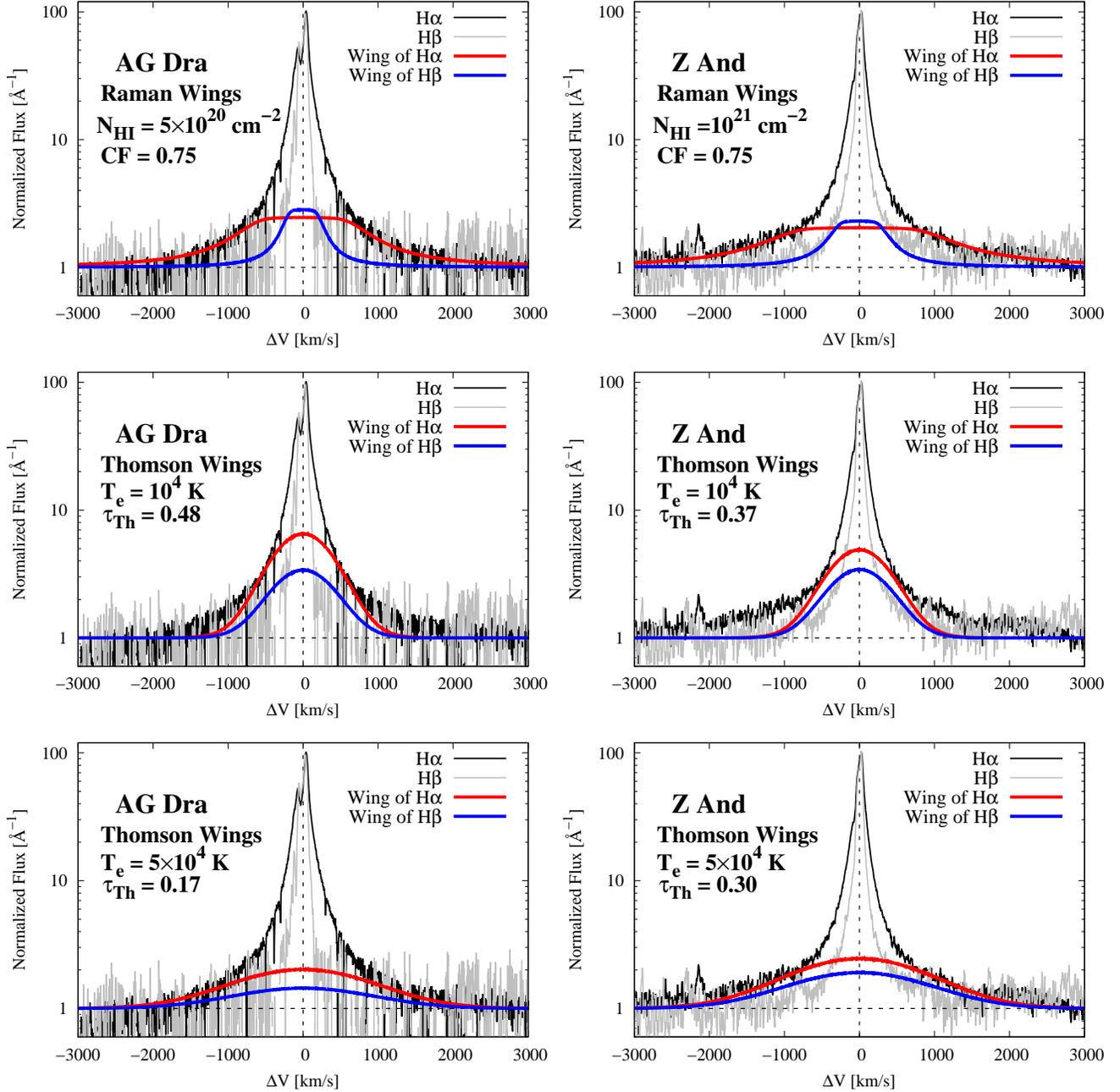}
\caption{Comparison of our Monte Carlo wing profiles and the {\it CFHT} spectrum around H$\alpha$ and H$\beta$ of AG Dra (left panels) and Z And (right panels) obtained 
with {\it CFHT}. The horizontal axis is the Doppler factor in units of ${\rm km\ s^{-1}}$.
The vertical scale is logarithmic normalized flux.
}
\label{obs_log}
\end{figure*}
In this subsection, we compare our simulated wing profiles with the observed data in the Doppler factor space.
H$\alpha$ and H$\beta$ profiles are normalized so that their peaks coincide with the numerical value of 101 and the underlying 
continuum has the value of 1. In Fig.~\ref{obs_log}, all the vertical axes are given in logarithmic scale.

The top two panels of Fig.~\ref{obs_log} show our simulated Raman H$\alpha$ and H$\beta$
wings in red and blue lines, respectively, which provide our best fit to the {\it CFHT} spectra
of AG~Dra and Z~And.  The best fit profiles are selected by eye.
The \ion{H}{1} column densities adopted for the simulation are $N_{HI} = 5\times 10^{20} \rm\ cm^{-2}$ and 
$10^{21} \rm \ cm^{-2}$ for
AG~Dra and Z~And, respectively, with $CF=0.75$ for both objects.
The contribution of the Raman wings to the observed H$\alpha$ and H$\beta$
is 15.5\% and 14.7\% for AG~Dra and 16.3\% and 12.2\% for Z~And.

We present a conversion between the relative flux scale in Fig.~\ref{cfht}
and physical scale for a typical symbiotic star characterized by the white dwarf mass 
 $M_*=0.6{\rm\ M_\odot}$ and a radius $R_*=0.01 {\rm\ R_\odot}$ \citep[][]{murset91}.
An accretion rate of $\dot M=10^{-7}{\rm\ M_\odot\  yr^{-1}}$  \citep[][]{tomov02,tomov08} gives rise to the accretion luminosity
\begin{equation}
L_{acc}={GM_* \dot M\over R_*} \sim 200\  L_\odot.
\end{equation}
According to \cite{murset91} the luminosities of AG~Dra and Z~And are $10\, L_\odot$ and $600\, L_\odot$, respectively,
whereas the temperature of the hot component is about $10^5{\rm\ K}$ for both the objects.

High ionization lines including \ion{O}{6}$\lambda\lambda$1032 and 1038 are common in many symbiotic stars indicating a hot
ionizing source with $T_h \sim 10^5{\rm\ K}$ \citep[][]{murset91,birriel00}.
 The specific luminosity at around Ly$\beta$ from a black body with a temperature $T_h=1.5\times 10^5{\rm\ K}$ and luminosity $L_{acc}$
 is
 \begin{equation}
 L_\lambda^{acc} \sim 8\times 10^{33}{\rm\ erg\ s^{-1}\ \AA^{-1}}.
 \label{acc_lum}
 \end{equation} 
The distances to AG~Dra and Z~And are poorly known and, for example, according to \cite{tomov02} and \cite{tomov08} they are 1.7 kpc for AG~Dra
and 1.1 kpc for Z~And, respectively. In terms of the distance $D$, the flux density $F_\lambda^{acc}$
near Ly$\beta$ associated with the accretion specific luminosity
in Eq.~(\ref{acc_lum}) is
 \begin{equation}
 F_\lambda^{acc} \sim 8.4\times 10^{-10} \left( {D\over 1 {\rm\ kpc}} \right)^{-2}{\rm\ erg\ cm^{-2}\ s^{-1}\ \AA^{-1}}.
 \label{acc_flux}
 \end{equation} 

In order to simulate the Raman wings around H$\alpha$ and H$\beta$ shown in the top panels of Fig.~\ref{obs_log},
continuum photons around Ly$\beta$ and Ly$\gamma$ are generated in the far~UV emission source in Fig~\ref{scheme}.
In our simulation, we consider the incident continuum flux ($ICF_{Ly\beta}$)
near Ly$\beta$ that is locally flat. The incident continuum flux is measured
in unit of $\rm \AA^{-1}$ for which the physical conversion of the unit value of
$ICF_{Ly\beta}=1{\rm\ \AA^{-1}}$ is discussed in Appendix B.

For the Raman wings of H$\alpha$ in Fig.~\ref{obs_log}, we find $ICF_{Ly\beta}=3.3\,\rm \AA^{-1}$ and $4.5\,\rm \AA^{-1}$ for AG~Dra and Z~And, respectively.
If the H$\alpha$ line flux is $2\times10^{-10}{\rm\ erg\ cm^{-2}\ s^{-1}}$, then these values are converted into
$ 2.7\times10^{-10}$ and $3.4\times10^{-10}\ {\rm\ erg\ cm^{-2}\ s^{-1}\ \AA^{-1}}$
for AG~Dra and Z~And, respectively.

A similar procedure is taken for $ICF_{Ly\gamma}$ near
Ly$\gamma$ for Raman wings around H$\beta$. In the simulations, we use $ICF_{Ly\gamma}=7.4\,\rm \AA^{-1}$ and $5.4\,\rm \AA^{-1}$ for AG~Dra and Z~And, respectively,
which are transformed to
$ 1.3\times10^{-11}$ and $2.0\times10^{-11}\ {\rm\ erg\ cm^{-2}\ s^{-1}\ \AA^{-1}}$, respectively. 
In Appendix B, the conversion of $ICF$ from our parametrization into physical units
is discussed.

The middle and bottom panels Fig.~\ref{obs_log} show Thomson wings simulated with two different values of $T_e=10^4{\rm\ K}$ and
$T_e=5\times 10^4{\rm\ K}$ for comparisons with the observed data.
In each panel, the Thomson H$\alpha$ profile
is stronger than Thomson H$\beta$ counterpart, simply because we injected more H$\alpha$ photons than H$\beta$ photons.
We compute the total number flux $TF_{H\alpha}$ of H$\alpha$ defined by
\begin{equation}
TF_{H\alpha} = \int_{\lambda(\Delta V=-500\,{\rm km/s})}^{\lambda(\Delta V = +500\,{\rm km/s})} NF_{H\alpha}(\lambda)d\lambda,
\end{equation}
where $NF_{H\alpha}(\lambda)$ is the observed flux normalized to have the peak value of 101. In a similar way, $TF_{H\beta}$ is computed to yield
the ratio of $TF_{H\alpha}/TF_{H\beta} = 2.0$ and 1.6 for AG~Dra
and Z~And, respectively.

These observed ratios are used to show Thomson H$\alpha$ and H$\beta$ wings in
Fig.~\ref{obs_log}.
In the middle panels ($T_e=10^4{\rm\ K}$), the Thomson optical depths are $\tau_{Th} = 0.48$ and 0.37 for AG~Dra and Z~And.
In the bottom panels ($T_e=5\times 10^4{\rm\ K}$),
 smaller values of $\tau_{Th}= 0.17$ and 0.30 are used for AG~Dra and Z~And, respectively.

In the middle panel, the near wing parts
 $|\Delta V|<500{\rm\ km\ s^{-1}}$  are well fit whereas the quality of fit becomes poor in the far wing parts. An opposite behavior is
 seen in the bottom panel, where good fit is obtained in the far wing parts. It should be noted that the electron temperature $T_e=5\times10^4
 {\rm\ K}$ is significantly higher than the values considered in the work of \cite{sekeras12} who analyzed the Thomson wing profiles of \ion{O}{6} and
 \ion{He}{2} emission lines.

\section{Conclusion and Discussion}

In this article, we used a Monte Carlo technique to produce broad wings 
around H$\alpha$ and H$\beta$. 
The adopted physical mechanisms include Thomson scattering of H$\alpha$ and H$\beta$ 
by free electrons and Raman scattering of Ly$\beta$ and Ly$\gamma$ by atomic hydrogen. 
In the case of Thomson wings, the wing profiles are well-approximated by a Gaussian function
whose width is mainly determined by the temperature of the free electron region. 
The Thomson scattering optical depth determines the resultant Thomson wing flux.

On the other hand, Raman wing profiles follow the Lorentzian function, which also describes 
the cross section. In the case of Raman wings, the wing flux is mainly affected 
by the \ion{H}{1} column density and the covering factor of the neutral region. The cross 
section near Ly$\beta$ is much larger than that near Ly$\gamma$, which leads to Raman 
H$\alpha$ wings much stronger than H$\beta$ counterpart.  When Balmer lines are overplotted 
in the Doppler factor space, Raman H$\alpha$ wings are expected to
appear stronger and broader than H$\beta$ wings.

We also presented high resolution spectra of the two S type symbiotic stars AG~Dra and Z~And 
obtained with the {\it CFHT} in 2014 in order to make quantitative comparisons with simulated 
Thomson and Raman wings.  The observational data and simulated data were shown in logarithmic 
scale, from which we notice that Raman wings provide better fit than Thomson wings. 


\cite{sekeras12} presented a convincing argument that the broad wings around \ion{O}{6} and
\ion{He}{2} lines are formed through Thomson scattering. It may be that Balmer emission lines are
formed in a more extended region than \ion{O}{6} and \ion{He}{2}, having higher
ionization potential than \ion{H}{1}. In this case, it is expected that the fraction of Thomson
scattered flux is much smaller for Balmer lines than for \ion{O}{6} and \ion{He}{2}. 

Considering the difficulty of explaining far wings with $\Delta V>10^3{\rm\ km\ s^{-1}}$ through Thomson scattering,
it remains as an interesting possibility that near wing parts are significantly contributed by Thomson
scattering whereas Raman scattering is the main contributor to the far wing parts.  
This possibility may imply complicated polarization behavior in the wing parts due to different scattering geometries
for Thomson and Raman processes.

Broad wings can also be formed from a hot tenuous wind around the compact component
with a high speed \citep[e.g.,][]{skopal06}. 
Because the flux ratio of H$\alpha$ and H$\beta$ in the fast wind will be
given by the recombination rate, it is expected that the flux ratio in the wing
part will be similar to that in the emission core part. This strongly implies
that H$\alpha$ and H$\beta$ will show similar profiles in the Doppler factor space.
In consideration of this, the disparity of wing widths in H$\alpha$ and H$\beta$
may indicate substantial contribution of Raman scattering to the Balmer wing formation.

Balmer wings formed through scattering are expected to develop significant linear 
polarization, implying that spectropolarimetry will shed much light on the physical origin. 
According to \cite{kim07}, Thomson wings may exhibit a complicated polarization structure 
because multiply scattered photons tend to contribute more to the far wing parts
than to the near wing parts. \cite{kim07} proposed that far wings 
are less strongly polarized than near wing regions. 

In the case of Raman scattering, wings may also be strongly polarized depending on the 
scattering geometry and the location of far UV emission source.  \cite{yoo02} investigated 
linear polarization and profiles of Raman scattered H$\alpha$ profiles. They proposed 
that the degree of linear polarization may increase as $\Delta V$ where single scattering 
dominates. \cite{ikeda04} carried out spectropolarimetry of AG~Dra and Z~And to propose 
that the polarization behavior of Raman O~VI$\lambda$ 6825  is similar to that of H$\alpha$ 
wings lending support to the Raman scattering origin. However, their analysis of H$\alpha$ 
was limited to a relatively narrow region $ |\Delta V|< 500{\rm\ km\ s^{-1}}$, 
which is insufficient to rule out the formation through Thomson scattering.

Broad H$\alpha$ wings are also found in young planetary nebulae including IC5117 and IC4997
\citep[e.g][]{hyung00, arrieta02}.
\cite{vandesteene00} also reported that some post-AGB stars show broad H$\alpha$ wings. 
The operation of Raman scattering by atomic hydrogen is also known for far UV \ion{He}{2}
lines near \ion{H}{1} Lyman series in symbiotic stars and a few planetary nebulae
including NGC~7027, NGC~6790, IC~5117
and NGC~6302 \citep[e.g.][]{kang09, lee06, pequignot97, groves02}.
Raman scattered \ion{He}{2} features are formed in the blue wing parts of Balmer emission lines.
Raman scattering cross section of \ion{He}{2}~$\lambda$1025 is
$\sigma\sim 7.2 \times 10^{21}{\rm\ cm^2}$, from which a broad scattered feature is formed at 6545 \AA.
The planetary nebulae with Raman \ion{He}{2} are known to exhibit broad H$\alpha$ wings,
which may be regarded as circumstantial evidence of contribution to Balmer wings
from Raman scattering.

\acknowledgments

We are grateful to the referee, who carefully read the manuscript and provided many constructive comments.
 This research was supported by the Korea Astronomy and Space Science Institute
under the R\&D program(Project No. 2018-1-860-00) supervised by the Ministry 
of Science, ICT and Future Planning.

\clearpage

\appendix

\section{Normalization of CFHT data}
\begin{figure*}
\centering
\includegraphics[scale=0.86]{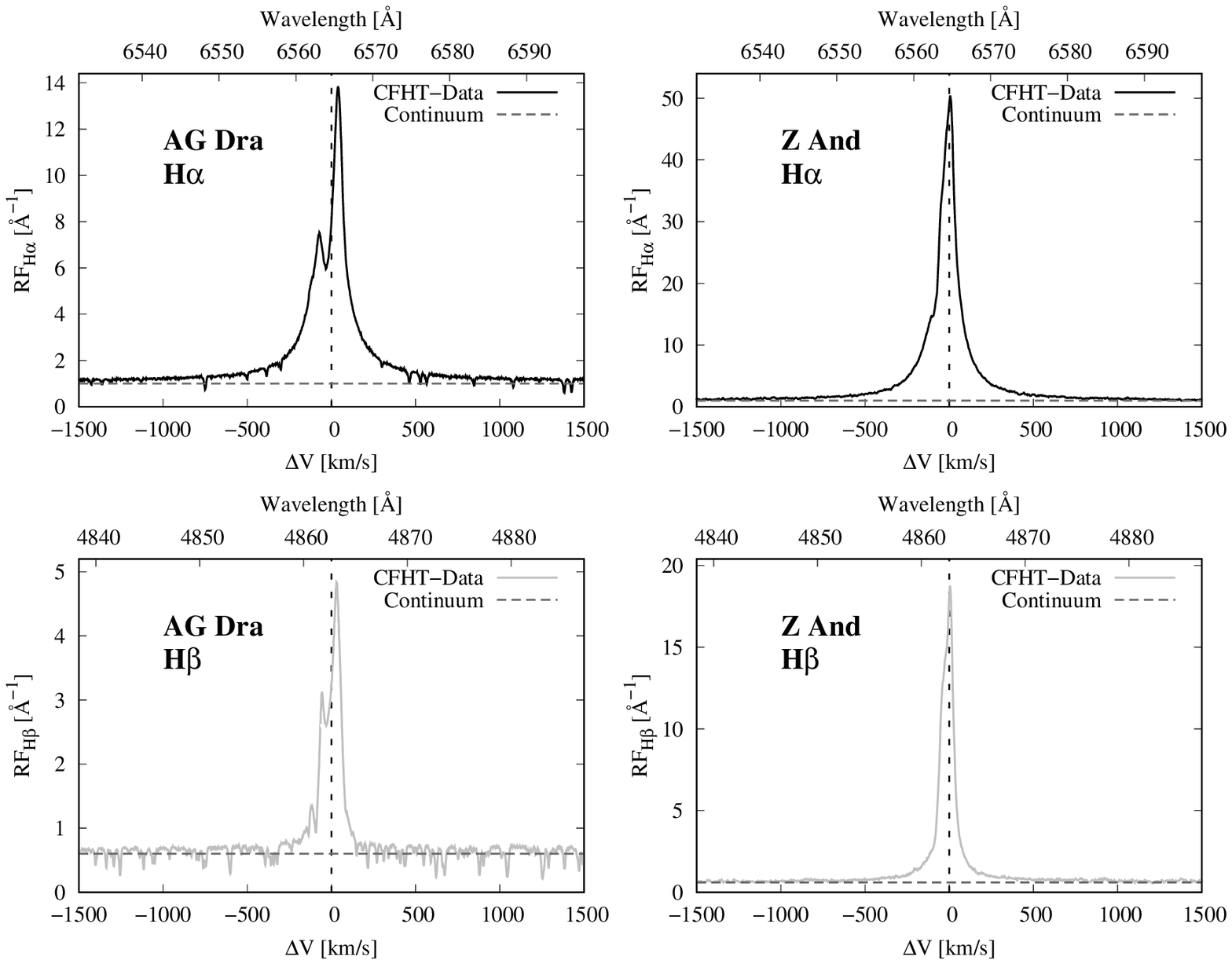}
\caption{Comparison of our Monte Carlo wing profiles and the {\it CFHT} spectra around H$\alpha$ and H$\beta$ of AG Dra (left panels) 
and Z And (right panels) obtained 
with {\it CFHT}.
The horizontal axises are the wavelength in units of $\rm\AA$ and the Doppler factor from line center in units of ${\rm km\ s^{-1}}$.
The vertical axis is the relative flux.
}
\label{obs_raw}
\end{figure*}

\begin{table*}[]
\centering
\caption{Parameters for the observed profiles of H$\alpha$ and H$\beta$ in Fig~\ref{obs_raw} }
\label{core}
\vskip5pt
\begin{tabular}{|l|l|c|c|c|c|c|}
\hline
Object& Line& $PV $ & $CL$ & $EW $  \\ 
\hline
AG~Dra&H$\alpha$ & 13.82 $\rm\AA^{-1}$  & 1.000 $\rm\AA^{-1}$  & 106.7 \AA  \\\hline
AG~Dra&H$\beta$  & 4.839 $\rm\AA^{-1}$  & 0.600 $\rm\AA^{-1}$  & 47.15 \AA   \\\hline
Z~And&H$\alpha$  & 50.44 $\rm\AA^{-1}$  & 1.000 $\rm\AA^{-1}$  & 238.8 \AA  \\\hline
Z~And&H$\beta$    & 18.77 $\rm\AA^{-1}$  &0.600 $\rm\AA^{-1}$  & 88.70 \AA   \\\hline
\end{tabular}
\label{convert}
\end{table*}

In Fig.~\ref{obs_raw}, we show the spectra of AG~Dra and Z~And obtained with the {\it CFHT}. The spectra
not being flux calibrated, the vertical axis shows the relative flux  ($RF$) in an arbitrary scale per \AA.
In turn,  Table~\ref{convert} shows the measured values of the peak value ($PV$), continuum level ($CL$) and equivalent width
($EW$) for the H$\alpha$ and H$\beta$ emission lines.   
Transformation of the relative flux into a physical
unit can be facilitated once the total line flux of an emission line is fixed. 
For example, \cite{tomov02} reported the total line flux of H$\alpha$
$F_{H\alpha}=2\times 10^{-10}{\rm\ erg\, cm^{-2}\, s^{-1}}$ for AG~Dra and a similar value was also reported for Z~And by \cite{tomov08}.
In this case, the unit relative flux ($RF=1 {\rm\ \AA^{-1}}$) in the spectra of Fig.~\ref{obs_raw} corresponds to
\begin{equation}
RF=1   {\rm\ \AA^{-1}}\,  \iff {F_{H\alpha}\over EW} \left({CL \over 1 {\rm\ \AA^{-1}}}\right)  =2 \times 10^{-12}
\left( {100\ {\rm\AA} \over EW} \right)
\left({CL \over 1{\rm\ \AA^{-1}}}\right)  {\rm\ erg\, cm^{-2}\, s^{-1}\, \AA^{-1}}. 
\end{equation}
Also it should be noted that the equivalent width $(EW)$ is independent of the scale.

In Figs.~\ref{cfht} and \ref{obs_log}, the normalized flux ($NF$) was introduced for profile comparisons of H$\alpha$ and
H$\beta$, where the normalized flux $(NF)$ is
related to $RF$ by the following relation 
\begin{equation}
NF = (RF-CL)\  \left[{{100 {\rm\ \AA^{-1}}} \over {PV-CL}} \right]+ (1{\rm\ \AA^{-1}}).
\end{equation}
With this relation, the normalized flux has the peak value to be 101 ${\rm\AA^{-1}}$ of $NF$  for both H$\alpha$ and H$\beta$. Because
$PV$ and $CL$ of H$\alpha$ differ from those of H$\beta$, different conversion into a physical unit is applied for
$NF_{H\alpha}$ around H$\alpha$ and $NF_{H\beta}$ around H$\beta$.

Around H$\alpha$, the unit value $NF_{H\alpha}=1{\rm\ \AA^{-1}}$ of the normalized flux corresponds to
\begin{equation}
NF_{H\alpha} = 1{\rm\ \AA^{-1}}\,
\iff \left( {F_{H\alpha}\over EW}\right) \ \left({CL \over 1{\rm\ \AA^{-1}}}\right) \ 
\left({PV-CL \over 100 {\rm\ \AA^{-1}}}\right),
\end{equation}
where $PV, CL$ and $EW$ are measured values for H$\alpha$.
With the fiducial value
of $F_{H\alpha}=2\times 10^{-10}{\rm\ erg\ s^{-1}\ cm^{-2}}$, we have
 \begin{equation}
NF_{H\alpha}=1   {\rm\ \AA^{-1}}\,
\iff 2 \times 10^{-12} \left[{PV-CL \over 100{\rm\ \AA^{-1}}} \right] \left[{CL\over 1{\rm\ \AA^{-1}}} \right] 
\left[{100{\rm\ \AA} \over EW}\right]   {\rm\ erg\, cm^{-2}\, s^{-1}\, \AA^{-1}}.
\label{halpha1}
\end{equation}

The ratios $F_{H\alpha}/F_{H\beta}$ measured from our {\it CFHT} spectra are 3.77 and 4.49 for AG~Dra and  Z~And, respectively.
Using these mesaured ratios, the normalized flux $NF_{H\beta}$ around H$\beta$ is converted into
\begin{equation}
NF_{H\beta} =1  {\rm\ \AA^{-1}}\,
\iff {F_{H\beta} \over EW}\ {CL \over 1{\rm\ \AA^{-1}}} \  {PV-CL \over 100{\rm\ \AA^{-1}}}.
\label{hbeta1}
\end{equation}
Adopting the ratio $F_{H\alpha}/F_{H\beta}$, $NF$ in unit of ${\rm\ \AA^{-1}}$ in the simulation is
converted to
\begin{equation}
NF_{H\beta} =1  {\rm\ \AA^{-1}}\,
\iff 2\times 10^{-12}\left[{F_{H\beta} \over F_{H\alpha}}\right]  
\left[{PV-CL\over 100{\rm\ \AA^{-1}}}\right] \left[ {CL\over 1{\rm\ \AA^{-1}}} \right] \left[ {100{\rm\ \AA}\over EW}\right]   
{\rm\ erg\, cm^{-2}\, s^{-1}\, \AA^{-1}}.
\label{hbeta2}
\end{equation}

\section{Converting from UV radiation to Raman scattering wing}
In our simulations, far UV continuum photons around Ly$\beta$ and Ly$\gamma$ are generated and
subsequently injected into neutral region to investigate the formation of Raman wings around H$\alpha$ and H$\beta$. 
Because the incident far UV continuum level in the simulation is also presented in unit of ${\rm\AA^{-1}}$ just like
the normalized flux around H$\alpha$ and H$\beta$ discussed above, we discuss the conversion of the
continuum level into the physical unit of ${\rm erg\ cm^{-2}\ s^{-1} \ \AA^{-1}}$.
 
Raman scattering relocates a far UV continuum photon into a Balmer wing region. The relocation involves the factor $(\lambda_o/\lambda_i)^2\simeq 40$
between the wavelength space near Ly$\beta$ and that near H$\alpha$. That is, a continuum interval of size $d\lambda_i$ around Ly$\beta$
is {\it ``Raman-transformed''} onto an interval around H$\alpha$ of size $d\lambda_o \simeq 40\ d\lambda_i$.  In addition, another factor of $\lambda_o/\lambda_i$
should be considered because an H$\alpha$ photon is less energetic than a Ly$\beta$ photon by the same factor.

Therefore,  the incident continuum flux ($ICF$) around Ly$\beta$ in unit of 
${\rm\AA^{-1}}$ (i.e., $ICF_{Ly\beta} = 1\ \rm \AA^{-1}$) is amplified by the factor $(\lambda_o/\lambda_i)^3\simeq 250$  
compared to $NF_{H\alpha}$ in Eq.~\ref{halpha1}.
In the case of $F_{H\alpha}=2\times 10^{-10}{\rm\ erg\ s^{-1}\ cm^{-2}}$,
the unit value of $ICF$ around Ly$\beta$ is converted into a physical unit of ${\rm\ erg\, cm^{-2}\, s^{-1}\, \AA^{-1}}$ as follows
\begin{equation}
ICF_{Ly\beta}=1{\rm\ \AA^{-1}}
\iff 5 \times 10^{-10} \left[{PV-CL \over 100{\rm\ \AA^{-1}}} \right] \left[{CL\over 1{\rm\ \AA^{-1}}} \right] 
\left[{100{\rm\ \AA} \over EW}\right]   {\rm\ erg\, cm^{-2}\, s^{-1}\, \AA^{-1}}.
\end{equation}
Here, $PV$, $CL$ and $EW$ are the parameters of the H$\alpha$ line presented in Table~\ref{convert}.

Similarly the factor of $(4850/972)^3\sim 100$ is involved in the wavelength
space near Ly$\gamma$ and that around H$\beta$.
Adopting Eq.~(\ref{hbeta2}), $ICF_{Ly\gamma}$ is
\begin{equation}
ICF_{Ly\gamma}=1{\rm\ \AA^{-1}}
\iff 2 \times 10^{-10} \left[{PV-CL \over 100{\rm\ \AA^{-1}}} \right] \left[{CL\over 1{\rm\ \AA^{-1}}} \right] 
\left[{100{\rm\ \AA} \over EW}\right]   {\rm\ erg\, cm^{-2}\, s^{-1}\, \AA^{-1}}.
\end{equation} 
The Raman wings around H$\alpha$ shown in Fig.~\ref{obs_log} are generated with the incident continuum flux 
$ICF_{Ly\beta}=3.3\, \rm \AA^{-1}$ and $4.5\, \rm \AA^{-1}$ for AG~Dra and Z~And, respectively. With the normalization of $F_{H\alpha}=2\times 10^{-10}{\rm\
erg\ cm^{-2}\ s^{-1}}$, the corresponding far UV continuum around Ly$\beta$ would be $2.7\times10^{-10}{\rm\ erg\ cm^{-2}\ s^{-1}\ \AA^{-1}}$
and $3.4\times  10^{-10}{\rm\ erg\ cm^{-2}\ s^{-1}\ \AA^{-1}}$ for AG~Dra and Z~And, respectively. Similarly for the Raman wings
around H$\beta$ shown in Fig.~\ref{obs_log}, $ICF_{Ly\gamma}=7.4\, \rm \AA^{-1}$ and $5.4\, \rm \AA^{-1}$ for AG~Dra and Z~And, respectively,
which are converted into $1.3 \times10^{-11}\ {\rm\ erg\ cm^{-2}\ s^{-1}\ \AA^{-1}}$ and $2.0\times10^{-11}\ {\rm\ erg\ cm^{-2}\ s^{-1}\ \AA^{-1}}$.


\end{document}